\documentclass[preprint,12pt]{elsarticle}

\usepackage{amsmath}
\usepackage{amssymb}
\usepackage{graphicx}
\usepackage{bm}
\usepackage{multirow}
\usepackage{hyperref}
\usepackage{color}
\usepackage{xcolor}
\usepackage{ulem}
\hypersetup{
  colorlinks = true,
  linkcolor = black,
  citecolor = blue,
  urlcolor = black,
  pdfauthor = {I G Tejada, L Brochard, T Leli\`evre, G Stoltz, F  Legoll and E Canc\`es},
  pdfkeywords = {molecular dynamics, multiscale modeling, graphene, rebo},
  pdftitle = {Coupling a reactive potential with a harmonic approximation for atomistic simulations of material failure},
  pdfsubject = {research article},
  pdfpagemode = UseNone
}

\journal{Computer Methods in Applied Mechanics and Engineering}

\begin{document}

\begin{frontmatter}



\title{Coupling a reactive potential with a harmonic approximation for atomistic simulations of material failure}


\author[navier]{Ignacio G. Tejada\corref{3sr}}
\ead{ignacio.tejada@3sr-grenoble.fr}
\author[navier]{Laurent Brochard}
\author[cermics,inria]{Tony Leli\`evre}
\author[cermics,inria]{Gabriel Stoltz}
\author[navier,inria]{Fr\'ed\'eric Legoll}
\author[cermics,inria]{Eric Canc\`es}


\cortext[3sr]{Corresponding author at: Universit\'e Grenoble Alpes, Laboratoire 3SR. Domaine Universitaire BP53 38041 Grenoble Cedex 9, France}

\address[navier]{Universit\'e Paris-Est, Laboratoire Navier (UMR 8205), CNRS, ENPC, IFSTTAR.  6 \& 8 avenue Blaise Pascal 77455 Marne-la-Vall\'ee, France.}
\address[cermics]{Universit\'e Paris-Est, CERMICS (ENPC), F-77455 Marne-la-Vall\'ee, France.}
\address[inria]{INRIA Rocquencourt, MATHERIALS project-team, 78153 Le Chesnay Cedex, France.}

\begin{abstract}
Molecular dynamics (MD) simulations involving reactive potentials can be used to model material failure. The empirical potentials which are used in such simulations are able to adapt to the atomic environment, at the expense of a significantly higher computational cost than non-reactive potentials. However, during a simulation of failure, the reactive ability is needed only in some limited parts of the system, where bonds break or form and the atomic environment changes. Therefore, simpler non-reactive potentials can be used in the remainder of the system, provided that such potentials reproduce correctly the behavior of the reactive potentials in this region, and that seamless coupling is ensured at the interface between the reactive and non-reactive regions. In this article, we propose a methodology to combine a reactive potential with a non-reactive approximation thereof, made of a set of harmonic pair and angle interactions and whose parameters are adjusted to predict the same energy, geometry and 
Hessian in the ground state of the potential. We present a methodology to construct the  non-reactive approximation of the reactive potential, and a way to couple these two potentials. We also propose a criterion for on-the-fly substitution of the reactive potential by its non-reactive approximation during a simulation. We illustrate the correctness of this hybrid technique for the case of MD simulation of failure in two-dimensional graphene originally modeled with REBO potential. 
\end{abstract}


\end{frontmatter}


\section{Introduction}

The atomic interactions and structure of a material are essential in determining its behavior and failure properties. Macroscopic failure mechanisms such as fracture or plastic deformation originate at the atomic scale, where events such as bond breaking and formation are the cause of cracks, vacancies and dislocation motions.

Atomistic approaches, which aim at relating the macroscopic description of a material to the underlying microscopic motion of  atoms, are appropriate to study the physics of mechanical failure. However, atomistic modeling of failure may be computationally expensive because systems of interest usually contain a very large number of particles (typically $\sim 100,000$ atoms) and because events at the atomic scale are particularly complex when approaching failure, and thus require elaborate numerical modeling that is expensive to use.
 
Among existing atomistic methods, quantum mechanics modeling provides the most fundamental approach but requires a high computational cost that limits its use to small systems and small time scales (typically less than 1000 atoms for a maximum of a few picoseconds). In contrast, classical molecular dynamics (MD) based on empirical potentials is an alternative technique that allows to reach system sizes and time scales relevant for the study of failure~\cite{Ashurst76, Abraham94, Abraham96}. It relies on the assumption that atoms move according to the laws of classical mechanics, where the force field is derived from an empirical potential representing the inter-atomic interactions. Accordingly, classical MD is not a `first principle' approach, and the results of the simulation depend on the choice of the energy potential. 

Therefore, the key ingredient in MD is the choice of the energy potential to model the mutual interactions of the atoms. This choice fully determines the physical relevance of the simulation. However it also has a significant impact on the computational cost of the simulation. Energy potentials are functions of the atom (or group of atoms) positions. Classical potentials are valid in the vicinity of a given atomic configuration, e.g., for solids in their elastic regime close to the reference configuration. In contrast, some potentials, called `reactive', have the ability to adapt the force field acting on an atom to its local environment. This extends their validity to a much wider range of configurations, e.g., mechanical failure of solids. Reactive potentials are usually calibrated on experiments and/or accurate quantum calculations~\cite{Abell85, Brenner90,Tersoff86, Stuart00, vanDuin01}. Most reactive potentials use a bond order formalism in which the energy potential is 
written as a sum of adaptive pairwise interactions that continuously adjust to the local environment. Since mechanical failure involves complex atomic rearrangements, such reactive potentials are often necessary in classical MD simulations of failure. In return of their adaptive capabilities, reactive potentials have a higher computational cost compared with simpler non-reactive potentials. For example, the computational costs of REBO~\cite{Brenner02}, Tersoff~\cite{Tersoff88}, AIREBO~\cite{Stuart00} and reaxFF~\cite{vanDuin01} potentials, as implemented in LAMMPS~\cite{Plimpton95}, are about $18$, $25$, $200$ and $680$ times that of a non reactive potential made of harmonic bonds, respectively, while the cost of a harmonic angle is about $3.5$ times that of a harmonic bond.\footnote{The origin of these ratios is twofold. The ratios between the performance of REBO, Tersoff, AIREBO and reaxFF are the results of the benchmark test provided with LAMMPS~\cite{Plimpton95} (http://lammps.
sandia.gov) for systems of $32,000$ atoms. The comparison of these performances with that of a harmonic bond potential comes from simulations of a face-centered cubic crystal with first neighbors bond interactions (12 neighbors per atom). No effect of system size was observed (systems of sizes between $500$ and $250,000$ atoms were considered). Note that the performance of a harmonic bond potential directly depends on the number of bonds in the system: a diamond crystal with $4$ first neighbors per atom would have a computational cost three times lower than the case considered in this example. Moreover, bonded interactions often include angular terms, the computational cost of which is about $3.5$ times that of bonds because of the angle computation, which is more expensive than the distance computation.} 

Atomic rearrangements during failure are often very localized and the adaptive capability of the potential is unnecessary elsewhere. Thus, substituting the reactive potential with a non-reactive analogue in all regions where reactivity 
is unneeded could significantly decrease the computational cost without affecting much the simulation results.

Following this idea, in this article, we propose a methodology to construct a hybrid MD simulation in which a reactive potential is coupled to a non-reactive analogue. The reactive potential is used where the reactive ability is needed and the non-reactive analogue is used elsewhere, where neither the stress nor the temperature are high, and the material behavior is elastic. In what follows, we will refer to the non-reactive analogue as the reduced potential, since it reproduces the reactive potential in a small region of the phase space close to a particular atomic topology and has no reactive ability.

The success of the hybrid approach depends on: (1) the ability of the reduced potential to correctly reproduce the reactive one when used; (2) the ability to avoid or circumvent nonphysical transition effects at the interface between the two potentials (seamless coupling). In this work, we develop an approach in which the reduced potential has a harmonic formulation that matches exactly the second order approximation of the reactive potential with respect to the positions of atoms in the ground state of the material (i.e., when the solid is unstressed and at 0 K). Doing so, we obtain a reduced potential that produces a dynamics identical to that of the reactive potential in the vicinity of the ground state, and our implementation of a seamless coupling is greatly facilitated for any combination of potentials in a hybrid  system. 

In the literature, similar hybrid techniques have been already proposed. For example, Buehler et al.~\cite{Buehler06} have coupled the reaxFF potential with the Tersoff potential to study dynamic crack propagation in a silicon single crystal. In contrast, we come up with a hybrid simulation technique that significantly differs from this approach (for instance we do not need a transition layer since the continuity of the energy and its derivatives is ensured by construction). We pay special attention to the validity of the coupling from a methodological point of view. In particular, we investigate the compatibility conditions for the two potentials and the appropriate way to make the transition from one potential to another. In a broader context, hybrid schemes in which quantum mechanics and molecular dynamics are combined have also been proposed~\cite{Bernstein09}, and they have some issues in common to MD/MD coupling, namely: whether or not the conservation of the energy is necessary, the conditions for a 
seamless coupling, \ldots

We consider pristine two-dimensional graphene as a case study to illustrate our methodology. Graphene is a material made of carbon atoms arranged on a two-dimensional honeycomb lattice. For simplicity, without loss of generality, we work in two spatial dimensions, i.e., out-of-plane bending and vibration modes are discarded. Although this affects the physical properties of graphene (in particular, phonon spectra and thermal expansion~\cite{Gao14}), this is not a shortcoming for the methodology itself.

The reactive potential we use to model pristine graphene is the 2nd-generation reactive empirical bond-order potential~\cite{Brenner02}, REBO. This potential accounts for the angular forces and gives the correct cohesive energy and equilibrium lattice constants of graphene, but is known to only imperfectly reproduce its elastic constants and phonon spectra~\cite{Tewary09}. This is a minor issue since we focus here on the coupling  methodology. 

In what follows, we describe the molecular dynamics model of pristine graphene in which either the reactive potential or its reduced version is used (Section~\ref{sc:ReducedPotential}). The methodology for coupling both potentials in hybrid models is explained in Section~\ref{sc:coupling}. We next propose a criterion to switch from one potential to the other along the simulation, in an on-the-fly manner (Section~\ref{sc:criterion}). Results obtained by using hybrid simulations of failure are presented and discussed in Section~\ref{sc:tests-dicussion}, followed by conclusions.

\section{Molecular model of pristine graphene}
\label{sc:ReducedPotential}

\subsection{Molecular dynamics}
\label{sbsc:MD}

We first describe the molecular dynamics of pristine graphene. The molecular system consists of a set of $N$  carbon atoms of mass $m$, whose spatial positions $\left\lbrace \bm{r}_\textit{1},\bm{r}_\textit{2}, \cdots, \bm{r}_N \right\rbrace$ evolve in time according to the laws of classical mechanics: 
\begin{equation}
\label{eq:newton}
\forall \, 1 \le I \le N \qquad m \bm{\ddot{r}}_I = - \nabla_I E_{(\bm{r}_\textit{1},\bm{r}_\textit{2}, \cdots, \bm{r}_N)} 
\end{equation}
\noindent where $E_{(\bm{r}_\textit{1},\bm{r}_\textit{2}, \cdots, \bm{r}_N)}$ is the inter-atomic energy potential, a function that depends on the positions of the atoms. The Langevin dynamics extends these equations to include friction and the stochastic effect of thermal agitation:
\begin{equation}
\label{eq:newton2}
\forall \, 1 \le I \le N \qquad m \bm{\ddot{r}}_I = - \nabla_I E_{(\bm{r}_\textit{1},\bm{r}_\textit{2}, \cdots, \bm{r}_N)} - \gamma m \bm{\dot{r}}_I + \sqrt{2 \gamma \text{k}_\text{B} T m} \bm{R}_{(t)}
\end{equation}
\noindent where $\gamma$ is the damping constant, $T$ is the temperature, $\text{k}_\text{B}$ is the Boltzmann's constant and $\bm{R}_{(t)}$ is a delta-correlated stationary Gaussian process with zero-mean, satisfying $\langle R_{\alpha (t)} \rangle = 0$ and $\langle R_{\alpha (t)} R_{\alpha (t')} \rangle = \delta_{\left( t - t'\right)}$ (here $\delta$ is the Dirac's delta and $\alpha$ represents the spatial coordinate).

The ground state, G.~S. , is the  configuration in which the total interaction energy reaches the absolute minimum.  In the case of graphene it corresponds to a honeycomb lattice of parameter $\sqrt{3} d$, where $d$ is the distance between the closest carbons (the value of $d$ depends on the potential used; here, it is of the order of $1.40$ \AA).

Various energy potentials can be used to model graphene. Here, we consider the REBO potential~\cite{Brenner02} which is a reactive potential developed to model any organic matter, including graphene. We describe here the formulation of Stuart et al.~\cite{Stuart00}. This potential is a reformulation of Brenner-Tersoff potentials~\cite{Brenner90,Tersoff86}, that are in turn based on Abell's formalism~\cite{Abell85} in which the binding energy is expressed as pairwise nearest-neighbor interactions that depend on the local atomic environment. The total REBO interaction energy is given by
\begin{equation}
\label{eq:reboenergy}
E^\text{REBO}_{(\bm{r}_\textit{1},\bm{r}_\textit{2}, \cdots, \bm{r}_N)}  = \sum_{I=1}^N  \sum_{J>I}^N E^{\text{REBO}, IJ}_{(\bm{r}_I, \bm{r}_J, \bm{r}_K, \cdots)}
\end{equation} 

\noindent with

\begin{equation}
\label{eq:rebobondenergy}
E^{\text{REBO}, IJ}_{(\bm{r}_\textit{I},\bm{r}_\textit{J}, \bm{r}_K, \cdots)}  = V^{IJ}_{\text{R}(r_{IJ}) }  - b^{IJ}_{(\bm{r}_I, \bm{r}_J, \bm{r}_{K}, \cdots)} \cdot V^{IJ}_{\text{A}(r_{IJ}) }
\end{equation}

\noindent where $V^{IJ}_{\text{R}}$ and $V^{IJ}_{\text{A}}$ are, respectively, the repulsive and attractive energies between atoms $I$ and $J$ (they only depend on the relative distance $r_{IJ}~=~\left\vert \bm{r}_J - \bm{r}_I \right\vert $  between atoms $I$ and $J$), $b^{IJ} $ is the so-called bond order parameter, in which not only $I$ and $J$ are involved but also other atoms $K$. REBO can be considered as a short range potential since only a limited number of nearby atoms determine the interaction energy of a bond. In particular, in two-dimensional graphene (in which dihedral angles do not play any role), only the first and second nearest neighbors of $I$ and $J$ determine the value of $b^{IJ}$. 

\subsection{Reduced potential}
\label{sbsc:reducedpot}
Our aim is to construct a reduced potential that mimics the REBO potential in the ground state of graphene (i.e., the ground states of both potentials share the same geometry and interaction energy) and in its vicinity.
Accordingly, the reduced potential should reproduce as much as possible the Taylor expansion in powers of displacements of atoms of REBO potential around the ground state.
The first order terms in the Taylor expansion  are the resulting forces on the atoms, which vanish in the ground state. The approach we propose consists in additionally
reproducing the second order term, which is given by the Hessian $\mathit{\mathit{\mathit{\Phi}}}$ of the potential energy $E$:

\begin{equation}
\label{eq:Hessian}
\mathit{\mathit{\mathit{\Phi}}}  =\left[ \mathit{\mathit{\mathit{\Phi}}}_{I \alpha J \beta}  \right] = \left[ \frac{\partial^2  E}{\partial r_{I \alpha} \partial r_{J \beta}} \right]
\end{equation}

\noindent where $I,J$ label atoms ($\textit{1}, \cdots, N$) and $\alpha,\beta$ represent the spatial coordinates ($x,y,z$). $\mathit{\mathit{\mathit{\Phi}}}$ is also called the force constant matrix, since under a harmonic approximation the total force in the direction $\alpha$ that atom $I$ experiences when atom $J$ is displaced of $u_{J \beta}$ from its position in the ground state configuration is given by $ \mathit{\mathit{\mathit{\Phi}}}_{I \alpha J \beta}  \cdot u_{J \beta} $. Many macroscopic properties of a material derive directly from the force constant matrix; in particular, the elasticity constants and the phonon dispersion curves, two critical properties for mechanics and fracture~\cite{Atrash11,Dove}. In short, the force constant matrix characterizes the energy of displacements at all length scales~\cite{Bernstein09}. Choosing a reduced potential that reproduces the Hessian ensures that the material properties essential to the study of failure are preserved. However, properties depending on 
higher order terms arising from REBO anharmonicity are 
not preserved, and therefore the proposed methodology is not suitable for the study of all material properties. We discuss these aspects in Section~\ref{sc:tests-dicussion}. 

To summarize, we seek a reduced potential $E^{\text{H}}$ that satisfies:
\begin{align}
\label{eq:conditions}
E^{\text{REBO} } \vert_\text{G.S.}  = E^{\text{H} } \vert_\text{G.S.}  = E_\text{G.S.} \nonumber \\
\left. \frac{\partial E^{\text{REBO} }}{\partial u_{I \alpha}} \right\vert_\text{G.S.} = 
\left. \frac{\partial E^{\text{H} }}{\partial u_{I \alpha}} \right\vert_\text{G.S.} = 0 \\
\left. \mathit{\mathit{\mathit{\Phi}}}^{\text{REBO}} \right\vert_\text{G.S.}  = \left. \mathit{\mathit{\mathit{\Phi}}}^{\text{H}} \right\vert_\text{G.S.}  = \mathit{\mathit{\mathit{\Phi}}}_\text{G.S.} \nonumber
\end{align}
\noindent where G.S. means that the configuration is that of the ground state (the same for both potentials).  
Many potentials $E^{\text{H}}$ meet these requirements. The most straightforward one is: 

\begin{equation}
\label{eq:reducedpot}
E^\text{H}  = E_\text{G.S.}  + \frac{1}{2} \sum_{I, J, \alpha, \beta} \mathit{\mathit{\mathit{\Phi}}}_{I \alpha J \beta} \cdot u_{I \alpha} \cdot u_{J \beta}
\end{equation}

Instead we decided to reproduce this potential in a constructive manner, by parametrizing a set of harmonic springs and angles connecting 
pairs and triplets of atoms, respectively. These elements (here referred to as `springs' and `angles') are widely implemented in molecular dynamics codes (in contrast to the potential~(\ref{eq:reducedpot})). This way of setting the potential 
up facilitates the on-the-fly construction or deconstruction and hence the potential substitution, whereas in a formulation like~(\ref{eq:reducedpot}) the 
contribution of each bond remains too implicit. We introduced this methodology previously in~\cite{Tejada15}. The potential energy is expressed as a quadratic 
function of the spring lengths $r_s$ and triplet angles $\theta_a$:

\begin{equation}
\label{eq:harmoenergy}
E^\text{H}  = E^\text{H}_0 + \sum_{s \in \text{springs}}  E^\text{S}_s + \sum_{a \in \text{angles}}  E^\text{A}_a
\end{equation} 

\noindent where 

\begin{align}
\label{eq:springsangles}
E^\text{S}_s = \frac{1}{2} K_s \left( r_s - r_{\text{eq},s} \right)^2 \\
E^\text{A}_a = \frac{1}{2} G_a \left( \theta_a - \theta_{\text{eq},a} \right)^2 \nonumber
\end{align}

\noindent One needs to find the topology of the springs and angles network and the values of the parameters $K_s$, $ r_{\text{eq},s} $,  $G_a$ and $\theta_{\text{eq},a}$ 
in order for Conditions~(\ref{eq:conditions}) to be satisfied.

First, we impose that all the springs and angles are at their minimum energies in the ground state (for any spring $r_{\text{eq},s}  = r^{\text{REBO}}_{\text{G.S.},s}  $
and for any angle $\theta_{\text{eq},a}  = \theta^{\text{REBO}}_{\text{G.S.},a} $) thus ensuring that the ground state of $E^\text{H}$ is the ground state of $E^{\text{REBO}}$. 
In view of Eqs.~(\ref{eq:harmoenergy})~and~(\ref{eq:springsangles}), the Hessian of the harmonic model can be written as
 
\begin{equation}
\label{eq:harmoHessian}
\left. \mathit{\mathit{\mathit{\Phi}}}^{\text{H}} \right\vert_\text{G.S.}  = \sum_{s \in \text{ springs}}  K_s \mathit{\mathit{\mathit{\Phi}}}^{\text{S}}_s + \sum_{a \in \text{angles}}  G_a  \mathit{\mathit{\mathit{\Phi}}}^{\text{A}}_a
\end{equation} 

\noindent where $\mathit{\mathit{\mathit{\Phi}}}^\text{S}_{s, I \alpha J \beta} = \left. \left( \frac{\partial r_s}{\partial r_{I \alpha} }\right) \left( \frac{\partial r_s}{\partial r_{J \beta} }\right) \right\vert_\text{G.S.}$ 
and $\mathit{\mathit{\mathit{\Phi}}}^\text{A}_{a, I \alpha J \beta} = \left. \left( \frac{\partial \theta_a}{ \partial r_{I \alpha} } \right) \left( \frac{\partial \theta_a}{\partial r_{J \beta} } \right)  \right\vert_\text{G.S.}$. 
$\mathit{\mathit{\mathit{\Phi}}}^\text{S}_{s}$ and $\mathit{\mathit{\mathit{\Phi}}}^\text{A}_{a}$ only depend on the atoms that are connected by the spring $s$ or the angle $a$ and on their positions in the ground state, 
that is on the given topology of springs and angles. $\left\lbrace \mathit{\mathit{\mathit{\Phi}}}^{\text{S}}_s, \mathit{\mathit{\mathit{\Phi}}}^{\text{A}}_a \right\rbrace$ is the basis of the subspace of Hessians that can 
be obtained with the given topology of springs and angles. 
 
By calibrating the values of $K_s$ and $G_a$ it is possible to approximate $\mathit{\mathit{\mathit{\Phi}}}^{\text{REBO}} \vert_\text{G.S.} $. We follow the algebraic method proposed by Mounet~\cite{Mounet05}
(used initially to enforce the acoustic sum rules and index symmetries on the Hessian of carbon allotropes obtained with density-functional perturbation theory). This method consists 
in projecting $\mathit{\mathit{\mathit{\Phi}}}^{\text{REBO}} \vert_\text{G.S.}$ onto the subspace defined by $\left\lbrace \mathit{\mathit{\mathit{\Phi}}}^{\text{S}}_s, \mathit{\mathit{\mathit{\Phi}}}^{\text{A}}_a \right\rbrace$. We consider the Frobenius
scalar product between two Hessian matrices $ \mathit{\mathit{\mathit{\Phi}}} \cdot \Psi = \sum_{I, \alpha, J, \beta} \mathit{\mathit{\mathit{\Phi}}}_{I \alpha J \beta} \Psi_{I \alpha J \beta}$. The distance associated to this scalar product
is $ \text{d} \left(\mathit{\mathit{\mathit{\Phi}}}, \Psi \right)~=~\sqrt{ (\mathit{\mathit{\mathit{\Phi}}}-\Psi) \cdot (\mathit{\mathit{\mathit{\Phi}}}-\Psi)} $. Finding the element $\mathit{\mathit{\mathit{\Phi}}}^{\text{H}} \vert_\text{G.S.}$ of the subspace which is the closest to the Hessian 
$\mathit{\mathit{\mathit{\Phi}}}^{\text{REBO}} \vert_\text{G.S.}$ is equivalent to projecting $\mathit{\mathit{\mathit{\Phi}}}^{\text{REBO}} \vert_\text{G.S.}$ onto the subspace. The basis $\left\lbrace \mathit{\mathit{\mathit{\Phi}}}^\text{S}_s, \mathit{\mathit{\mathit{\Phi}}}^\text{A}_a \right\rbrace$ can be orthonormalized with the scalar product previously defined to obtain a new basis $\left\lbrace U_o \right\rbrace$ 
in which, for any $o$ and $p$,  $\text{d} \left(U_o, U_p \right) = \delta^o_p$ (being $\delta^o_p$ the Kronecker symbol). Then the best approximation $\mathit{\mathit{\mathit{\Phi}}}^\text{H} \vert_\text{G.S.}$ is 
obtained as the projection of $\mathit{\mathit{\mathit{\Phi}}}^\text{R} \vert_\text{G.S.} $ onto the new basis, i.e., $\mathit{\mathit{\mathit{\Phi}}}^\text{H} \vert_\text{G.S.}~=~\sum_o~(\mathit{\mathit{\mathit{\Phi}}}^\text{R} \vert_\text{G.S.}~\cdot~U_o~)~U_o$. 
Finally $\mathit{\mathit{\mathit{\Phi}}}^\text{H} \vert_\text{G.S.}$ can be expressed in the original basis. We thus obtain the values of the stiffness constants $K_s$ and $G_a$.  Furthermore, 
when $ \text{d} \left(\mathit{\mathit{\mathit{\Phi}}}^\text{R} \vert_\text{G.S.} , \mathit{\mathit{\mathit{\Phi}}}^\text{H} \vert_\text{G.S.} \right) = 0$, then the prescribed topology of springs and angles is able to exactly reproduce
the original force constant matrix.

This methodology can be applied to any atomistic model, but when the system is a crystal (like graphene), then some symmetries exist, and as a consequence many of the angle and spring parameters
are identical by symmetry. Producing a Hessian consistent with these material symmetries is precisely the aim of the method described in~\cite{Mounet05}.

\begin{figure}[t]
\begin{center}
\includegraphics[width=0.4\textwidth]{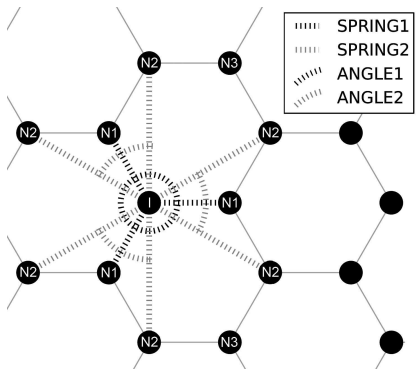}
\caption{\label{fig:schematic} Schematic representation of the topology of springs and angles that connect an atom $I$ to its first ($N1$) and second ($N2$) nearest neighbors.}
\end{center}
\end{figure}

In~\cite{Tejada15} we have followed this procedure to construct a reduced version of the REBO potential valid for pristine two-dimensional graphene. With four classes of harmonic interactions, as shown in 
Figure~\ref{fig:schematic} and made precise in Table~\ref{tab:approx}, our approximated model reproduces exactly the geometry, potential energy and Hessian of the 
reactive model in the ground state. Therefore, when atoms experience small displacements from their equilibrium positions in the ground state, this reduced potential accurately approximates 
the reactive potential. 

\begin{table}
\caption{\label{tab:approx} Parameters of the springs and angles used in the approximated model (Figure~\ref{fig:schematic})}
{\footnotesize
\begin{center}
\begin{tabular}{llll}
\multirow{2}{*}{Interaction} & \multirow{2}{*}{Atoms connected} & \multirow{2}{*}{Stiffness constant} & Equilibrium \\
& & & length or angle \\
Spring 1 & $I$-$N1$ & $K_{1}=39.86$ eV/ \AA$^2$ & $r_{\text{eq},1}=1.398$ \AA \\
Spring 2 & $I$-$N2$  & $K_2=1.29$ eV/ \AA$^2$ & $r_{\text{eq},2}=2.421$ \AA \\
Angle 1 & $N1$-$I$-$N1$ & $G_1=1.33$ eV & $\theta_{\text{eq},1} = 2 \pi / 3$ \\
Angle 2 & $N2$-$I$-$N2$ & $G_2=6.87$ eV & $\theta_{\text{eq},2} = \pi / 3$\\
\end{tabular}
\end{center}
}
\end{table}

\begin{figure}[t]
\begin{center}
\includegraphics[width=0.5\textwidth]{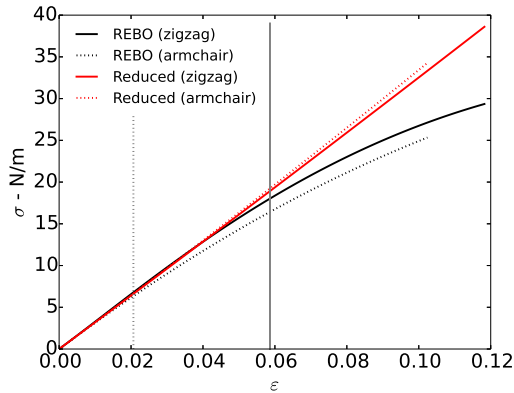}
\caption{\label{fig:bulkmechanic} Comparison of the tensile response of 2D graphene in armchair and zigzag directions obtained from MD simulation with REBO and the reduced potential shown in Figure~\ref{fig:schematic} 
and made precise in Table~\ref{tab:approx}. The vertical lines represent the strains at which the relative error in stress reaches $5\%$. The response of the reduced potential is a straight line in the $(\varepsilon, \sigma)$ diagram.}
\end{center}
\end{figure}

To check the quality of the approximation near the ground state and to quantify the effects of anharmonicity as the system moves away from that state, we compare the tensile response of bulk graphene obtained with REBO and with the reduced potential (see Figure~\ref{fig:bulkmechanic}). 

The crystalline structure was stretched in zigzag and armchair directions while the respective perpendicular directions were left undeformed. For each deformed state the energy was minimized and the stress computed. We used LAMMPS~\cite{Plimpton95} to carry out these numerical experiments. These tests perfectly illustrate how the approximation fails as the strain increases. However they just reveal the error in the forces (but not in the energy) of two of the many cases of lattice deformation that are present in a failure test. Therefore an adaptative criterion to exchange the potentials must be set up taking into account more situations of lattice deformation (Section~\ref{sc:criterion}).

Another way of assessing the importance of anharmonicity is to use lattice dynamics and the so-called quasi-harmonic approximation~\cite{Dove} to observe how the phonon spectrum changes
with temperature. Note that there is no thermal expansion with harmonic models and that at 0~K the harmonic and the reduced potentials produce exactly the same phonon spectra since 
they have exactly the same Hessian. However, because of the small thermal expansion obtained for two-dimensional graphene (about $10^{-6}$ K$^{-1}$~\cite{Gao14}), we checked that only at 
very high temperatures the lattice dynamics differs from one potential to the other (see our previous work~\cite{Tejada15}). For example, the calculated frequency at the K point 
differs by less than $2$ THz for a temperature increase of $1800$ K, while the wave group velocities hardly changed.

\section{Coupling methodology}
\label{sc:coupling}

In this section we describe the methodology to couple the reactive and reduced potentials in a hybrid MD simulation.

\subsection{Transition between potentials}
\label{sbsc:transition}
A possible strategy to couple the reactive and reduced potentials is to use a transition and a buffer zone at the interface between the reactive and the non-reactive regions (see e.g. Buehler et al.~\cite{Buehler06}). This is not the strategy we will follow but we describe it because it is a common choice. In this approach, both potentials are computed in the transition layer and weighted with a smooth interpolation 
function to obtain the total interaction energy. One could also imagine interpolating the force instead of the energy. Results a priori depend on the size of the transition region. The buffer zone is an extension of the transition zone that is needed to ensure that 
potentials are properly computed in the transition zone. Such a smooth transition ensures the continuity and differentiability of the hybrid potential, but raises questions from a
methodological point of view. In particular, any shift of energy between the reduced and reactive potentials (because the former is only an approximation of the latter) leads
to artificial forces in the transition layer since the weighting procedure introduces a gradient of energy orthogonal to the transition layer. 
Conversely, if forces are interpolated instead of energy, then there is no global underlying energy.

Instead of a transition zone, the coupling methodology we develop here consists in an abrupt transition for which seamless coupling is ensured by considering a particular set of harmonic springs and correction forces at the interface between the potentials.

Before proceeding further, we want to mention that similar problems arise for methods coupling reference atomistic models with coarse-grained atomistic models involving less degrees of freedom. The QuasiContinuum Method (QCM) is one such methods~\cite{Tadmor96, Tadmor03}. Several variants have been proposed. Some of them use a transition zone in which both models are concurrently taken into account through a weighting function. Some other variants are based on an abrupt transition between both models. The coupling can be either performed in terms of energy (in which case a hybrid energy is defined for the whole system) or in terms of forces (in which case forces on each atom are defined in a hybrid way, and there is no underlying energy for the whole system). We refer to~\cite{Luskin13,Miller09} for a review on these variants.

\begin{figure}[t]
\begin{center}
\includegraphics[width=0.5\textwidth]{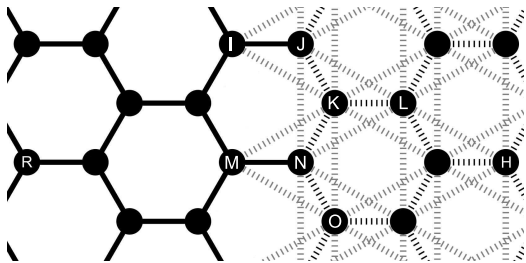}
\caption{\label{fig:hybrid} Schematic representation of the interface between reactive and non-reactive zones. Solid lines represent reactive bonds and dotted lines represent springs (for clarity, angles are not depicted).}
\end{center}
\end{figure}

We now detail our methodology. Within a hybrid model, we denote as a reactive bond  
the interaction between a pair of nearest neighbors $I$ and $J$  as formulated with the bond order formalism of Eq.~(\ref{eq:rebobondenergy}) (the bond is precisely labeled
by atoms $I$ and $J$). Note that other atoms take part in this expression through the $b^{IJ}$ coefficient. We will define the reactive zone $\mathcal{R}$ as the set of reactive bonds
and the non-reactive zone $\mathcal{H}$ as the set of harmonic elements. In the bulk of $\mathcal{R}$ and $\mathcal{H}$ zones (e.g. for the atoms $R$ and 
$H$ in Figure~\ref{fig:hybrid}), interactions are modeled, respectively, with Eq.~(\ref{eq:rebobondenergy})
or with the Eqs.~(\ref{eq:springsangles}) used to construct the reduced potential. 

A special treatment is needed near the interface. For instance, in Figure~\ref{fig:hybrid}, the constant of the spring
connecting atoms $I$ and $K$ is not clearly defined since the bond $I$ - $J$ is reactive while the bond $J$ - $K$ is not. Indeed, with the topology and values of springs and angles as they are set in the
previous section, no obvious correspondence exists between a particular reactive bond in Eq.~(\ref{eq:rebobondenergy}) and a particular spring / angle in Eq.~(\ref{eq:harmoenergy}).
Each harmonic element does not exclusively substitute a single reactive bond. Therefore the set of springs and angles near the interface  has to be determined separately in 
order to ensure no redundant or missing energy between the reactive and reduced potentials. 
 
To obtain the particular values of the harmonic elements near the transition interface, the methodology used in the previous section to construct the reduced potential can be adapted. The energy of a hybrid system can be written as

\begin{equation}
E^\text{HYB}_{(\bm{r}_\textit{1},\bm{r}_\textit{2}, \cdots, \bm{r}_N)}  =  E^\text{HYB,REBO}_{(\bm{r}_\textit{I} \in \mathcal{R})} + E^\text{HYB,H}_{(\bm{r}_\textit{I} \in \mathcal{H})}
\end{equation}

\noindent with $E^\text{HYB}_{(\bm{r}_\textit{1},\bm{r}_\textit{2}, \cdots, \bm{r}_N)} $ representing the total energy of the hybrid system and
$  E^\text{HYB,REBO}_{(\bm{r}_\textit{I} \in \mathcal{R})} $ and  $ E^\text{HYB,H}_{(\bm{r}_\textit{I} \in \mathcal{H})}$ representing the contributions of the reactive 
and reduced potentials, respectively.
Accordingly, the Hessian is the superposition of the corresponding Hessians. We can write: $\mathit{\mathit{\mathit{\Phi}}}^\text{HYB} = \mathit{\mathit{\mathit{\Phi}}}^\text{HYB,REBO} + \mathit{\mathit{\mathit{\Phi}}}^\text{HYB,H}$. In the ground state
$\mathit{\mathit{\mathit{\Phi}}}^\text{HYB,H} \vert_\text{G.S.} = \sum_{s \in \text{ springs}}  K_s \mathit{\mathit{\mathit{\Phi}}}^\text{S}_s + \sum_{a \in \text{angles}}  G_a  \mathit{\mathit{\mathit{\Phi}}}^\text{A}_a$. To ensure that the hybrid system
approximates the reactive potential, the springs and angles constants should be chosen in order to satisfy: $\mathit{\mathit{\mathit{\Phi}}}^\text{HYB} \vert_\text{G.S.} = \mathit{\mathit{\mathit{\Phi}}}^\text{REBO} \vert_\text{G.S.}$. 
We do so by projecting $\mathit{\mathit{\mathit{\Phi}}}^\text{REBO} \vert_\text{G.S.}  - \mathit{\mathit{\mathit{\Phi}}}^\text{HYB,REBO} \vert_\text{G.S.}$ onto the subspace of possible $\mathit{\mathit{\mathit{\Phi}}}^\text{HYB,H} \vert_\text{G.S.}$. 
In the present case, the chosen topology of springs and angles allows to fully capture the reactive potential (i.e. $\mathit{\mathit{\mathit{\Phi}}}^\text{HYB}$ is exactly equal to $\mathit{\mathit{\mathit{\Phi}}}^\text{REBO}$).  This approach can be applied to any transition interface,
 but must be repeated each time the interface is modified to enlarge or reduce the reactive zone. In practice, one could identify and characterize all elementary interface
 geometries so as to treat any interface as the superposition of elementary interfaces. But doing so would be cumbersome. Instead, we opted for a much easier 
 way to determine the springs and angles values at the interfaces, as explained hereafter in Subsection~\ref{sbsc:bondperbond}.

\subsection{Bond per bond substitution}
\label{sbsc:bondperbond}

\begin{figure}[t]
\begin{center}
\includegraphics[width=0.7\textwidth]{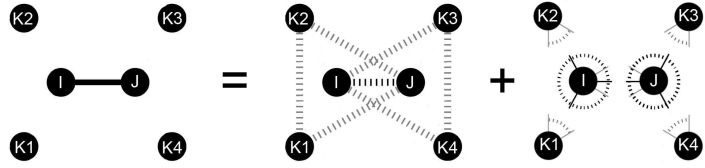}
\caption{\label{fig:bondperbond} Schematic representation of the bond per bond substitution approach. A reactive bond $I$-$J$ (whose energy also depends on atoms $K1$ to $K4$) can be replaced by
a system made of seven springs and 12 angles, as made precise in Table~\ref{tab:bondperbond}.}
\end{center}
\end{figure}

The bond order formalism, often used by reactive potentials including REBO, formulates the total energy as a sum over pairs of atoms, which we have referred to as reactive bonds.
We take advantage of this formalism to propose a `bond per bond' substitution technique to determine the particular springs and angles constants in the vicinity of the transition interface. 
In practice, we  can determine the exact set of springs and angles that approximates the energy of a single REBO bond between two atoms $I$ and $J$ within a full reactive 
environment $E^{\text{REBO}}_{IJ(\bm{r}_I, \bm{r}_J, \bm{r}_K, \cdots)}$. Indeed, we know the contribution $\mathit{\mathit{\mathit{\Phi}}}^{\text{REBO}}_{IJ}$ of this single bond to the 
total Hessian in the ground state (because of Eqs.~(\ref{eq:reboenergy})~and~(\ref{eq:Hessian})), such that $\mathit{\mathit{\mathit{\Phi}}}^\text{REBO} = \sum_I \sum_{J>I} \mathit{\mathit{\mathit{\Phi}}}^{\text{REBO}}_{IJ}$. Thus, we can 
easily determine the set of strings and angles that reproduces this partial Hessian $\mathit{\mathit{\mathit{\Phi}}}^{\text{REBO}}_{IJ}$. In the case of REBO simulation of 
graphene, each 
reactive bond can be replaced by a set of 7 springs and 12 angles as shown in Figure~\ref{fig:bondperbond} and made precise in Table~\ref{tab:bondperbond}. We 
followed the same methodology explained above for arbitrary transition interfaces to obtain the values of the various parameters. In what follows, we refer to the 
set of springs and angles approximating a single reactive bond $I$-$J$ as `harmonic bond $I$-$J$'.

\begin{table}
\caption{\label{tab:bondperbond} Parameters of the springs and angles used in a bond per bond substitution approach (see Figure~\ref{fig:bondperbond})}
{\footnotesize
\begin{center}
\begin{tabular}{llll}
\multirow{2}{*}{Interaction} & \multirow{2}{*}{Atoms connected} & \multirow{2}{*}{Stiffness constant} & Equilibrium \\
& & & length or angle \\
Spring 1 & $I$-$J$ & $K_{1}=39.86$ eV/ \AA$^2$ & $r_{\text{eq},1}=1.398$ \AA \\
Springs 2+ & $I$-$K3$, $I$-$K4$, $J$-$K1$, $J$-$K2$  & $K_2^+=1.34$ eV/ \AA$^2$ & $r^+_{\text{eq},2}=2.4208$ \AA \\
Springs 2-- & $K1$-$K2$, $K3$-$K4$  & $K_2^-=-1.39$ eV/ \AA$^2$ & $r^-_{\text{eq},2}=2.421$ \AA \\
\multirow{2}{*}{Angles 1+} & $J$-$I$-$K1$, $J$-$I$-$K2$, & \multirow{2}{*}{$G_1^+=1.78$ eV} & \multirow{2}{*}{$\theta^+_{\text{eq},1} = 2 \pi / 3$} \\
 & $I$-$J$-$K3$, $I$-$J$-$K4$ & & \\
Angles 1-- & $K1$-$I$-$K2$, $K3$-$J$-$K4$ & $G_1^-=-2.22$ eV & $\theta^-_{\text{eq},1} = 2 \pi / 3$ \\
Angles 2+ & $K3$-$I$-$K4$, $K1$-$J$-$K2$ & $G_2^+=11.51$ eV & $\theta^+_{\text{eq},2} = \pi / 3$\\
\multirow{2}{*}{Angles 2--} & $J$-$K1$-$K2$, $J$-$K2$-$K1$, & \multirow{2}{*}{$G_2^-=-2.32$ eV} & \multirow{2}{*}{$\theta^-_{\text{eq},2} = \pi / 3$}\\
& $I$-$K3$-$K4$, $I$-$K4$-$K3$ & & \\
\end{tabular}
\end{center}
}
\end{table}

To set the hybrid system, we carry out a bond per bond substitution. Indeed, the springs and angles stiffness constants can be split into the contributions of each bond, irrespective of whether an atom is in bulk graphene or near the interface. Each reactive bond removed on the reactive potential side adds its harmonic counterpart 
to the network of springs and angles. In the bulk of the harmonic zone, these contributions add up exactly to the values of springs and angles constants obtained in 
Section~\ref{sc:ReducedPotential}. In the vicinity of the transition interface, the total spring and angle constants differ from those values. Most importantly this methodology 
preserves the Hessian through the transition interface. As an illustration, the constant of the spring connecting atoms $J$ and $N$ in Figure~\ref{fig:hybrid}, $K_{J-N}$, can 
be seen as the superposition of the constants of the corresponding springs involved in the substitution of  three reactive bonds ($J$ - $K$,
 $K$ - $L$ 
and $K$ - $N$), so $K_{J-N} = K_2 = K^+_2+K^-_2+K^+_2$ (where $K^+_2$ and $K^-_2$ are given in Table~\ref{tab:bondperbond}, and $K_2$ in Table~\ref{tab:approx}). Similarly, we find that 
$K_{K-M} = K^+_2+K^-_2 \neq K_2$ since the reactive bond $M$-$N$ is left reactive.

The main advantage of the bond per bond substitution approach is that it easily adapts  to any interface  at almost no computational cost, thus allowing for fast on-the-fly modification of the reactive zone during a simulation. However, there is a remaining issue to be taken care of when the bond per bond substitution approach is used, namely the equilibrium of forces on the atoms at the transition interface.  We explain the issue and the way we handle it in Section~\ref{sbsc:interfaceforces}.

\subsection{Interface forces}
\label{sbsc:interfaceforces}

The constructed reduced potential has been set up to model systems with the same geometry, equilibrium energy and Hessian in the ground state as systems described by a full reactive potential. With both potentials, the system in the ground state is in equilibrium, i.e. the total forces acting on the atoms are equal to zero. However, the partial forces exerted by each bond on any atom are not necessarily zero. In contrast, by construction, all the forces exerted by springs and angles on the corresponding atoms are zero for the reduced harmonic potential; this is due to the fact that the equilibrium distances and angles were set a priori to the lattice distances and angles in the ground state.

\begin{figure}[t]
\begin{center}
\includegraphics[width=0.8\textwidth]{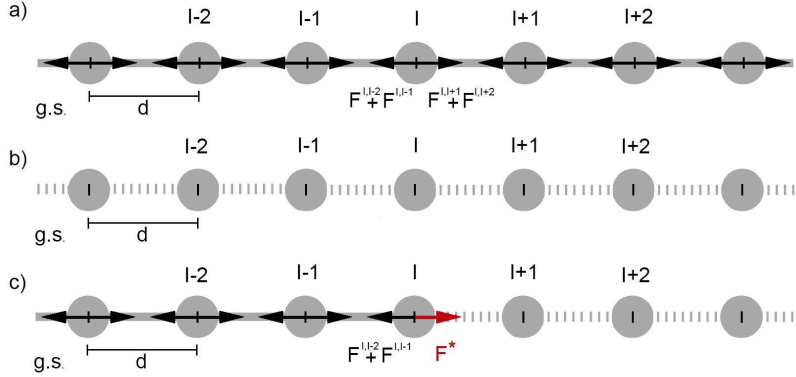}
\caption{\label{fig:interface} Illutrative example of a one-dimensional system in its ground state modeled with: (a) a reactive potential, (b) a reduced potential and (c) a hybrid scheme.}
\end{center}
\end{figure}

Therefore, both models keep atoms balanced but the underlying forces are different. This difference has no implication when  considering bulk graphene entirely modeled with the reactive or the non-reactive potential, but a problem arises for hybrid systems: at the interface between reactive and reduced potentials, atoms are not properly equilibrated since the forces caused by the reactive bonds near the interface are not completely canceled out by the set of springs and angles that make up the reduced potential. Formally, the Taylor expansion of the hybrid energy potential in the ground state exhibits non-zero first order terms, corresponding to the derivatives of the potential with respect to the interface atoms positions. This issue can be illustrated by the schematic example depicted in Figure~\ref{fig:interface}. The same issue arises in the QCM method mentioned in Section~\ref{sbsc:transition}.

A strategy to solve this issue consists in adding constant balancing forces on the interface atoms in order to leave all the atoms balanced when the system is in the ground state. Equivalently, these balancing forces can be interpreted as new terms in the potential energy of the hybrid system, proportional to the displacement of the interface atoms in the direction opposite to the force. Since the second derivatives of these terms are zero, they do not affect the Hessian of the hybrid system. However, they contribute to the first order terms in the Taylor expansion of the hybrid energy potential, ensuring zero total values of the forces. Therefore, these balancing forces correct the spurious interface forces while preserving the second order approximation. A similar approach has been used in some QCM approaches, where these additional forces are referred to as ghost forces. In practice, the balancing forces can be included at almost no computational cost during bond by bond substitution. The constant values 
of the forces produced by a single reactive bond with the geometry of the ground state are computed just once. Then, when at a given time in the simulation an atom $I$ is at an interface (i.e. when some but not all the bonds in which it participates are reactive~\footnote{Note that in the cases of reactive potentials that involve distant neighbor interactions, external forces must be added not only on the atoms right at the interface between reactive and harmonic zones but also on each atom that is affected by the substitution of a reactive bond.}), the corresponding forces, properly oriented, are added to each of the harmonic bonds. The forces produced by the reactive bonds together with these artifitial forces leave the atom force-balanced. 

\section{Substitution criterion}
\label{sc:criterion}
When the configuration of the system of atoms is near the ground state, the reduced potential is an excellent approximation of the reactive potential. Conversely, the approximation fails when the explored configurations differ significantly from the ground state, because of excessive deformation or temperature. Therefore, we need a practical  criterion quantifying the quality of the approximation and that can be used to trigger the substitution of the reactive potential by the reduced potential and vice-versa. Moreover, on-the-fly adaptive coupling requires a simple analytic criterion with minimum marginal computational cost. 

\begin{figure}[t]
\begin{center}
\includegraphics[width=0.30\textwidth]{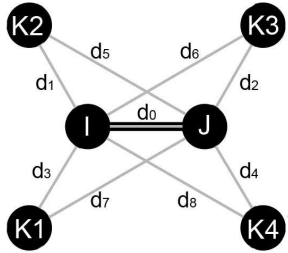}
\caption{\label{fig:distances} Schematic representation of a reactive bond $I$-$J$ with the atoms involved and the distances that fully characterize the environment.}
\end{center}
\end{figure}

We propose to construct a criterion based on the 9 distances that fully characterize the environment of a reactive bond, as shown in Figure~\ref{fig:distances} (this environment is made of 6 atoms, i.e. 9 degrees of freedom in 2D after removing the rigid body motion). These distances are computed at each timestep to obtain the harmonic forces. The computation of the criterion has therefore a very limited additional computational cost. The primary objective of the criterion is to quantify the mismatch between the reduced potential and the reactive potential in such a way that  values of the criterion exceeding a threshold value characterize appropriately excessive  errors that make the substitution of potentials  unsuitable. There is however no simple relationship between the geometry and the energy or the forces. Accordingly, a geometry-based criterion is necessarily empirical. We evaluated different formulations of the criterion by assessing its relation to the error on a large variety of deformed 
configurations. In the present case, we considered the error in the 
energy only, but one could also consider the error for some other quantities such as the forces or the phonon frequencies of the crystal (i.e. the eigenvalues of the Hessian).

Along with the expression of the criterion, one has to determine a threshold value. An optimal choice of the threshold is essential to ensure the quality of the approximation at a minimum computational cost: if the threshold is too small, bonds turn reactive very far from failure, and the efficiency of the simulation will hardly be improved; if the threshold is too large, the error induced by the substitution of potentials causes excessive deviations from the reference material behavior  and spurious dynamics effects (e.g., nonphysical heating or cooling because of energy gaps at substitution).

As mentioned above, a reactive bond in 2D graphene modeled with REBO involves 6 atoms. We constructed a criterion based on 9 distances that fully characterize the relative positions of those 6 atoms: the 5 nearest-neighbor distances ($d_0$ to $d_4$) and 4 of the second-order distances ($d_5$ to $d_8$) (see Figure~\ref{fig:distances}). As a starting point, we consider as a criterion the sum over the 9 segments of the relative differences between the distance in the current configuration and that in the ground state. However, we observed that the criterion is slightly improved (i.e. it produces a better correlation between the error in the energy and the criterion) when only one of the nearest-neighbor distances, $d_0$, is considered. Therefore the criterion we use is finally written as: 

\begin{equation}
\label{eq:criterion}
C = \frac{1}{5} \left( C^{IJ} + C^{IK_3} + C^{IK_4} + C^{JK_1} + C^{JK_2} \right)
\end{equation}
\noindent with
\begin{equation}
C^{ij} =  \frac{\vert r^{ij} - r^{ij}_\text{G.S.} \vert}{r^{ij}_\text{G.S.}}
\end{equation}

\begin{figure}[t]
\begin{center}
\includegraphics[width=0.40\textwidth]{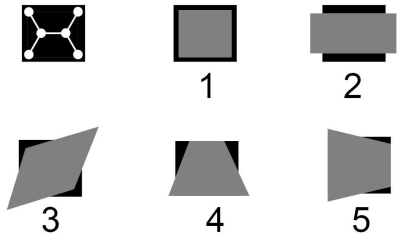}
\caption{\label{fig:deformation} Deformation modes of a two-dimensional rectangle.}
\end{center}
\end{figure}

\begin{figure}[t]
\begin{center}
\includegraphics[width=0.7\textwidth]{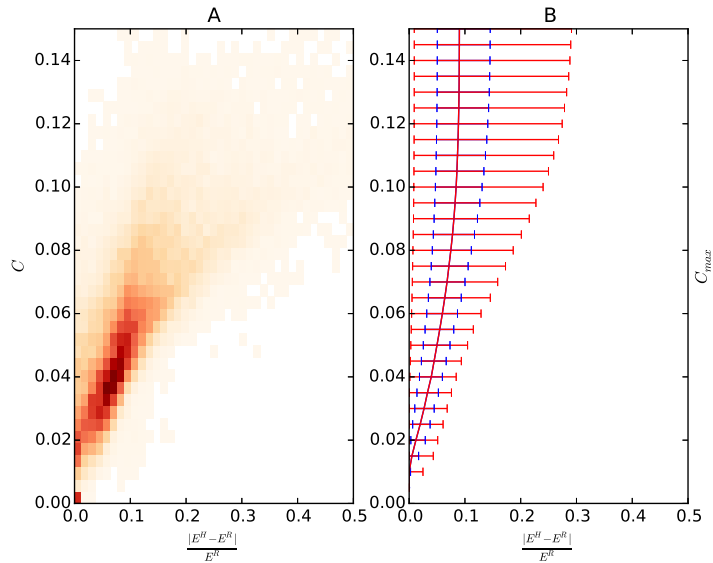}
\caption{\label{fig:criterion} A sample of deformed configurations was generated. For each deformed configuration, the values of the criterion (Eq.~(\ref{eq:criterion})) and of the error in the energy ($\frac{\vert E^\text{H} - E^\text{R} \vert}{E^\text{R}} $) were computed. A: 2D histogram. B: Box plots of the values of the error of all the configurations whose criterion lies below a given value $C_\text{max}$. The blue whiskers represent  the first and third quartiles, the continuous line is the median and the leftmost / rightmost whisker represents the 5$^\text{th}$ / 95$^\text{th}$ percentile.}
\end{center}
\end{figure}

We tested this criterion on a set of deformed configurations to assess the relation between the criterion and the error in energy for a single reactive bond. The set of configurations was generated as follows: we impose some deformations and then follow a Metropolis algorithm~\cite{Metropolis53} to sample configurations according to Boltzmann's statistics at finite temperature, i.e., the temperature of interest for application of the methodology. Here, we consider 300~K. The specific deformations we impose correspond to the two-dimensional deformation modes of a rectangle: isotropic, deviatoric, distortion and two bending modes (see Figure~\ref{fig:deformation}). The amplitude of these deformations was increased linearly (in positive and negative directions for modes 1 and 2) and, for each deformed configuration, 100 configurations were generated with the Metropolis algorithm. We used a uniform jumping distribution, $r_{I,\alpha}~\rightarrow~r_{I,\alpha}~+~\Delta r_{I,\alpha}$ with $\Delta r_{I,\alpha}~\sim~\
mathcal{U}(-\delta,\delta)$ and $\delta=0.04$~\AA, obtaining an average acceptance ratio of around $0.5$. We expect this set of configurations to be representative of the configurations encountered in actual simulations of failure, thus reducing the  bias in the 
testing of the criterion. Results are shown in Figure~\ref{fig:criterion}.

With this chart, one can readily determine the appropriate criterion threshold corresponding to a given level of error. An ideal criterion would characterize the error exactly with no uncertainty. Dispersion to the right of the mean corresponds to configurations with low values of the criterion but large errors which is a risk for the reliability of the coupling. Dispersion to the left of the mean corresponds to configurations with large values of the criterion but small errors which is a potential cause of inefficiency of the coupling. Therefore, an appropriate criterion should keep the dispersion of the chart as small as possible. In fact this allows to quantify the relevance of a given criterion and opens the way for a more systematic investigation.

\section{Graphene failure test cases and discussion}
\label{sc:tests-dicussion}

As a test case of the proposed methodology, we performed simulation of failure of the 2D pristine graphene. We considered two different cases of fracture: tensile tests on samples with a prescribed initial crack and tensile tests on samples with a preexisting hole. Because of stress concentrations in the vicinity of the initiated crack or hole, one expects the reactive zones to grow from there. Stress concentrations around cracks and holes are quite different, thus the two test cases complement each other.

We used an in-house implementation of the coupling methodology. The program LAMMPS~\cite{Plimpton95} was used as a benchmark to validate our implementation of classical MD and of the REBO potential. The value of the threshold for the criterion was set by trial and error. Periodic boundary conditions were imposed. For simulations with crack, the crack was initiated by ignoring some bonds along the zigzag direction. For simulations with hole, the hole was initiated by removing all atoms lying inside a circle. All the atoms were initially non-reactive, except those located where the crystalline structure ends (at the boundaries of the hole) or around the initially broken bonds.

\begin{figure}[t]
\begin{center}
\includegraphics[width=0.7\textwidth]{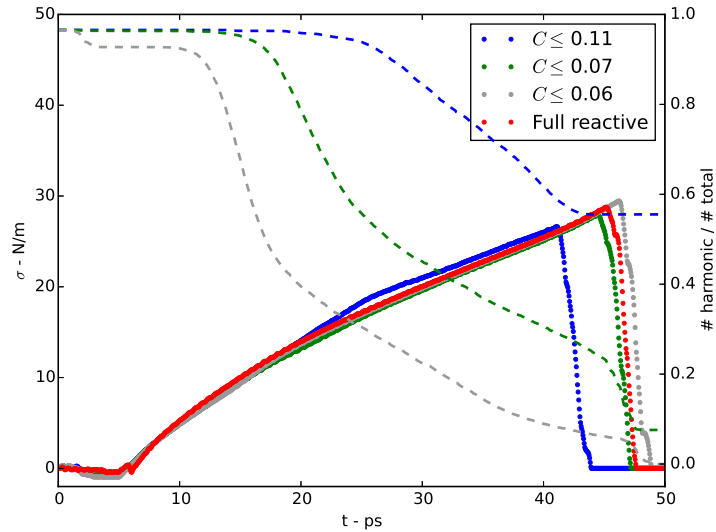}
\caption{\label{fig:crack} Average mechanical response of the precracked system computed with the hybrid methodology when  different thresholds are imposed to the substitution criterion. The dotted lines, associated to the right-hand axis,  represent the fraction of harmonic atoms at each time.}
\end{center}
\end{figure}

A Langevin thermostat was used everywhere to impose a temperature of $300$~K. Once proper thermal equilibrium is reached, a mechanical loading is applied while the thermostat is kept. The systems were deformed at a constant strain rate until failure. The deformation was imposed  in the armchair direction, while the system was left undeformed in the perpendicular direction. Atoms were not remapped during the deformation (i.e. only the dimensions of the simulation box were changed but the positions of atoms are not adjusted to the new box but are left to move according to the dynamics). A strain rate of $0.5$ \% / ps was used which is small enough to avoid rate dependent behaviors. Therefore, we follow a quasi-static evolution of the system from the ground state until failure. Post failure evolution is disregarded here since the choice of boundary conditions and thermostat is unsuitable for the study of crack propagation in brittle solids.  More details of these simulations are given in  \ref{ap:simulations}.

\begin{figure}[t]
\begin{center}
\includegraphics[width=0.7\textwidth]{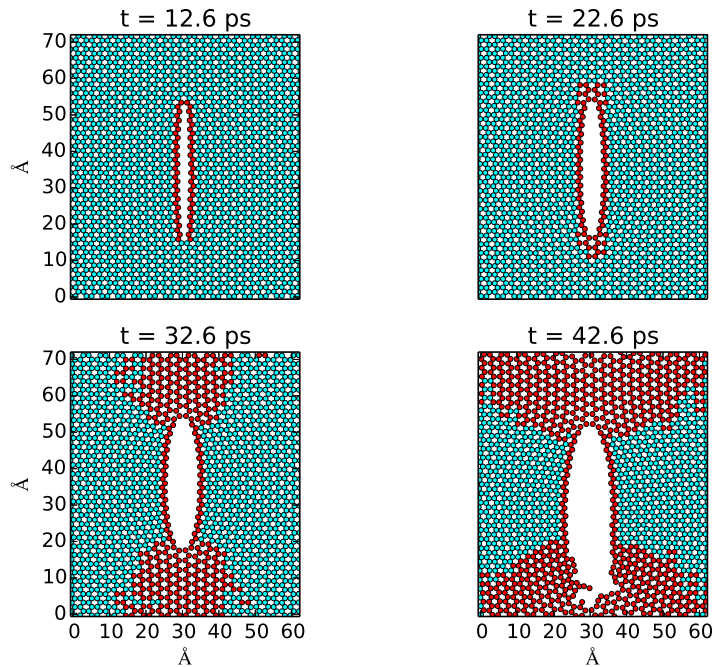}
\caption{\label{fig:cracksnapshots} Snapshots of graphene fracture in a MD simulation with the hybrid methodology we propose (with $C \leq 0.11$) and a system with a preexisting crack. Atoms in red are those connected with the reactive potential.}
\end{center}
\end{figure}

\begin{figure}[t]
\begin{center}
\includegraphics[width=0.7\textwidth]{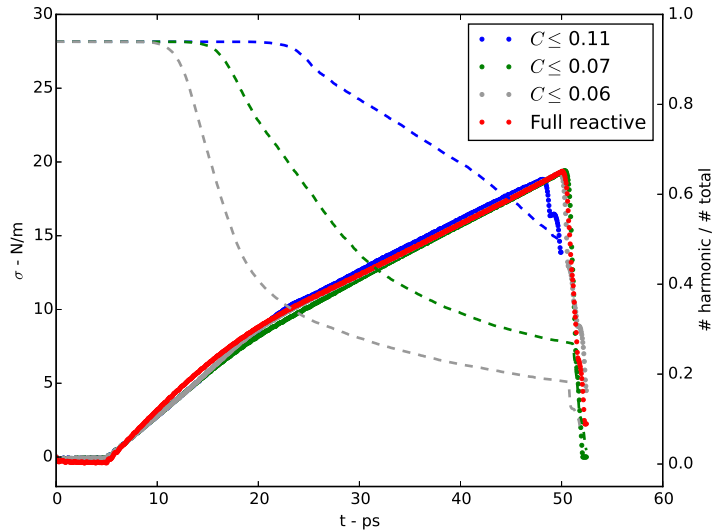}
\caption{\label{fig:hole} Average mechanical response of the system with a hole computed with the hybrid methodology we propose when  different thresholds are imposed to the substitution criterion. The dotted lines, associated to the right-hand axis,  represent the fraction of harmonic atoms at each time.}
\end{center}
\end{figure}

\begin{figure}[t]
\begin{center}
\includegraphics[width=0.7\textwidth]{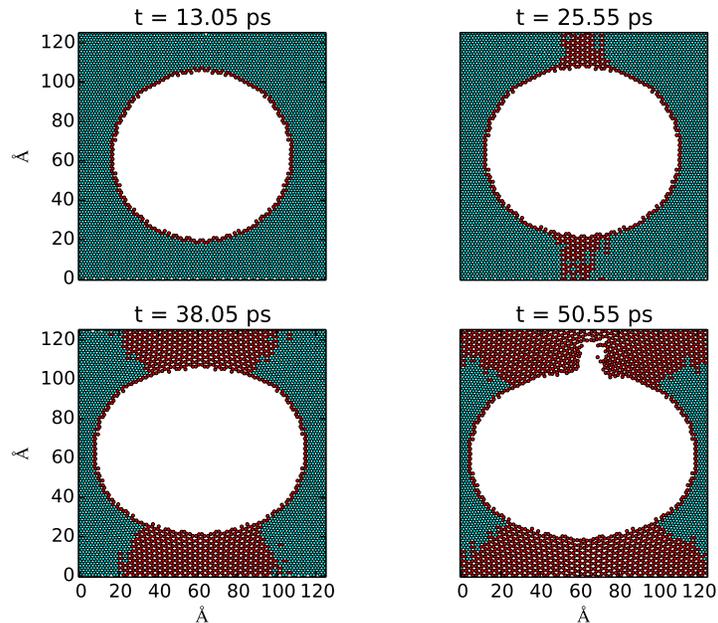}
\caption{\label{fig:holesnapshots} Snapshots of graphene fracture in a MD simulation with the hybrid methodology we propose (with $C \leq 0.11$) and a system with a preexisting hole. Atoms in red are those connected with the reactive potential.}
\end{center}
\end{figure}

We compare in Figure~\ref{fig:crack} the average mechanical response (average over ten different realizations for each threshold) obtained when the system with preexisting crack is simulated with the reactive potential only and when it is simulated with the proposed coupling methodology. The average mechanical response compares well over the whole loading process until failure appears. The mechanical behaviors start to be significantly different only with the highest threshold. Figure~\ref{fig:cracksnapshots} displays some snapshots of the system during one of the simulations.  Likewise, Figures~\ref{fig:hole} and~\ref{fig:holesnapshots} display the average mechanical responses for the system with a preexisting hole and a few snapshots of the corresponding molecular configurations, respectively. Again, the results of the hybrid potential accurately reproduce those of the reactive potential over the whole loading range. In this test differences from one threshold to another were less significant. 

In view of these results we can say that the coupling methodology successfully reproduces the expected behavior of these two examples of 2D graphene failure.  As expected, the size of the reactive zone increases as the system is stretched and extends in the direction of stress concentration. 

The efficiency of the coupling methodology strongly depends on the fraction of the system that is simulated with the reactive potential. Obviously, if most of the bonds are reactive, no significant computational improvement can be expected from the proposed methodology. Since the coupling methodology adds only modest computational cost (calculation and potential substitution are all computation free, and the computation of interface forces only requires to identify interface atoms), a reasonable  estimation of the ratio of the computational time of a hybrid simulation to that of its fully reactive equivalent  is simply given by 

\begin{equation}
\frac{t^\text{HYB}}{t^\text{R}} = \left\langle r \right\rangle + \frac{c^\text{H}}{c^\text{R}}(1 - \left\langle r \right\rangle)
\end{equation}

\noindent where $c^\text{H}$ and $c^\text{R}$ are the computational costs (seconds per atom and timestep) of the harmonic and reactive potentials, respectively, and $\left\langle r \right\rangle = \frac{1}{n_\text{steps}} \sum_i^{n_\text{steps}} \frac{N^\text{R}_i}{N}$ is the average ratio of the number of reactive atoms $N^\text{R}$ to the total number of atoms $N$. 

The computational cost of the reduced potential is low because its implementation relies on lists of neighbors, i.e., the neighboring environment is fixed. In contrast, the reactive potential recomputes the neighboring environment at each timestep. We estimated the ratio $c^\text{R}/c^\text{H}$ to be approximately $5.2$ in simulations with LAMMPS~\cite{Plimpton95} on systems of two-dimensional graphene of $10^3$, $10^4$ and $10^5$ atoms (no system size effect was observed). This ratio is rather consistent with the benchmark mentioned in the introduction\footnote{In introduction, we reported that REBO's computational cost is 18 times larger than that of a potential made of 6 harmonic bonds per atom (face-centered cubic crystal with 12 neighbors). In addition the computational cost of a harmonic angle is about 3.5 times larger than that of a harmonic bond. Therefore, for the reduced potential of 2D graphene which includes 4.5 harmonic bonds and 6 angles per atom, one could anticipate an efficiency $\frac{1}{4.
5 / \left(18 \cdot 6\right) + 6 \cdot 3.5 / \left(18 \cdot 6\right)} = 4.2$ times better than REBO.}.

The values of $\left\langle r \right\rangle$ can vary widely from one system to another. Systems in which stresses are highly concentrated in a small region have $\left\langle r \right\rangle \ll 1$ and can strongly benefit from the coupling ($\frac{t^\text{HYB}}{t^\text{R}} \approx \frac{c^\text{H}}{c^\text{R}}$), whereas systems with very little stress concentration have $\left\langle r \right\rangle \approx 1$ so that the coupling is ineffective ($\frac{t^\text{HYB}}{t^\text{R}} \approx 1$). In the case studies presented above for validation purposes, stress is not so concentrated because of the small size of the system, hence the rather large fraction of reactive atoms (see the values of $\left\langle r \right\rangle$ reported in Table~\ref{tab:averagereactive}). For these test cases the computation time of the simulations may be reduced by a factor of three with the coupling methodology.

\begin{table}
\caption{\label{tab:averagereactive} Values of $\left\langle r \right\rangle$ (average over the whole loading curve) resulting from the hybrid simulations of fracture for the different threshold values of the criterion.}
{\footnotesize
\begin{center}
\begin{tabular}{ccc}
Threshold & Precracked system &  System with a hole\\
$C_\text{max}$ & $\left\langle r \right\rangle$ &   $\left\langle r \right\rangle$\\
$0.06$ & $0.64$ & $0.55$\\
$0.07$ & $0.46$ & $0.45$\\
$0.11$ & $0.21$ & $0.19$\\
\end{tabular}
\end{center}
}
\end{table}

The coupling methodology suffers from some limitations that we discuss hereafter. First, there is an inherent trade-off between efficiency and accuracy. Indeed, in the case of 2D pristine graphene modeled with REBO we were able to propose a reduced potential that matches exactly the Hessian of the reactive potential. If instead, we had selected a less complex reduced potential, e.g., without second neighbors springs, with a lower computational cost, this reduced potential would have only approximated the Hessian of the reactive potential. In general, reproducing exactly the Hessian of a reactive potential may require quite complex reduced potentials, especially if the reactive potential involves distant neighbor interactions. It is then tempting to consider simpler reduced potential at the expense of the ability of this potential to accurately reproduce the behavior of the material. For purely static behaviors (e.g., mechanical elasticity at low temperature), reduced potentials limited to close neighbors 
interactions are in principle sufficient. However, for more acute features (e.g., vibrational modes), accounting for distant neighbors interactions may prove to be critical. Therefore, one faces an inherent trade-off between accuracy and efficiency when applying the proposed methodology. 

A second limitation of the approach is the second order approximation used. Indeed, various material properties, including thermal expansion and conductivity, non-linear mechanics, etc., depend on the anharmonicity of the potential, i.e., on higher order terms in the Taylor expansion.  Since the reduced potential is constructed to only reproduce the second order expansion of the potential, those properties are not captured by the reduced potential. As a consequence, the proposed methodology is unable to estimate such properties. In addition, this is a source of inefficiency of the proposed methodology, since the substitution of potentials is valid only as long as anharmonic terms are small compared to harmonic terms (small strain and temperature). Therefore, a reactive potential with significant anharmonicity is substitutable only at low strain and temperature. The only way to circumvent this issue would be to include some anharmonicity effects within the reduced potential (e.g. by using non harmonic springs 
and angles).

A last issue is related to on-the-fly substitution. On-the-fly substitution means that reactive bonds are substituted (or recreated) during the simulation. The differences in energy and forces between the reduced potential and the reactive one lead to artificial jumps in energy and forces during that process. Spurious behaviors may occur. For instance, if the same reactive bond is substituted and recreated periodically, significant amounts of energy could be artificially pumped in or out of the system. In the present case, the thermostat equilibrates the system as the loading proceeds and maintains the energy. From a practical point of view, the substitution criterion is critical for this issue. Substitution should occur when the error is small. The criterion should also avoid leaving isolated harmonic bonds inside the reactive regions and vice-versa.

\section{Conclusion and perspectives}
 
In this article, we present a methodology to speed up molecular dynamics simulations involving reactive potentials and, as an illustrative test case, we apply the proposed methodology to the simulation of failure of two dimensional graphene. This methodology consists in setting up hybrid models where reactive potentials are substituted with reduced non-reactive potentials. The reduced potential approximating a reactive potential is constructed from a set of harmonic bond and angle interactions, allowing one to reproduce the behavior of the system close to the ground state as predicted by the reactive potential (same geometry, same interaction energy and same Hessian). The values of the parameters to use in the harmonic interactions can be obtained by an algebraic methodology. The reduced potential is a second order approximation of a reactive potential which ensures a correct lattice dynamics  around the ground state. This approximation holds as long as higher order terms in the 
Taylor expansion of the reactive potential around the ground state are negligible. Non negligible contributions may be due to high deformation or high temperature.

The coupling of the reactive and reduced potentials in a hybrid simulation is performed through an abrupt transition between reactive and non-reactive regions. The seamless coupling is ensured by considering a particular set of harmonic springs and correction forces at the interface. For potentials based on the bond order formalism, this particular set can be easily adapted during a simulation, according to the specific reactive bonds that are being replaced at each time. We deal with the issue of extra forces needed to equilibrate the forces on the atoms at the interface between reactive and non-reactive regions. We solve this problem by adding constant forces, which affect neither the energy nor the Hessian. 

On-the-fly adaptive coupling requires a criterion to automatically switch from harmonic to reactive interactions during a simulation. We propose a criterion based on the geometrical differences between the considered configuration and the ground state configuration. The precise expression of the criterion is calibrated by assessing its correlation with errors in energy over a set of plausible deformed states sampled according to Boltzmann's statistics at finite temperature. Finally we determine the criterion threshold triggering substitution by trial and error in hybrid MD simulations.

We illustrate the various aspects of this methodology with a hybrid simulation of fracture in graphene. The reactive potential was REBO and the reduced version is made of a set of springs and angles connecting each atom to its first and second order neighbors. The results of this test case satisfactorily reproduce the expected fully reactive behavior and illustrate the interest of the approach. The coupling methodology is efficient if only a small fraction of the system is simulated with the reactive potential, that is, if stresses and reactive processes are highly concentrated. In the example of the reduced potential proposed for 2D graphene the coupling can decrease the computational cost by a factor of 3.

Some issues remain open to further research with the goal of improving the performance of simulations without compromising the validity of the results. In particular, the criterion for on-the-fly substitution can be more deeply studied to reduce the fraction of the system that is computed with reactive potentials at a given simulation time. It is also interesting to analyze whether or not a simpler version of reduced potential could be used far from the reactive zone, or how, in contrast, construct 
a reduced version including some anharmonic effects. In the end, a multi-scale hybrid approach in which reactive potentials and different reduced versions are coupled can  make an important difference in the performance of molecular dynamics simulation of failure but it requires a careful consideration of the physical implications, especially of those relating to the seamless coupling. Finally, the methodology presented here could be adapted to long-range reactive potentials.

\section*{Acknowledgments}
This work has benefited from a French government grant managed by ANR within the frame of the national program Investments for the Future ANR-11-LABX-022-01.

 \appendix

 \section{Details of hybrid MD simulations}
 \label{ap:simulations}
{\small
 \begin{itemize}
	\item Features of the samples of pristine graphene:
 	\begin{itemize}
 	 	\item Number of atoms: $N=1,800$ (precracked) and $N=4,062$ (hole)
 	 	\item Initial configuration: Ground state G.S. (honeycomb; $a=\sqrt{3} d$)
		\item Distance between closest atoms in the G.S.: $d = 1.3977$ \AA
		\item Potential energy per bond in the G.S.: $e_\text{G.S.} = \frac{2 E_\text{G.S.}}{3 N} = -5.2049$~eV
		\item Partial forces exerted by a single bond on each atom in the ground state (according to the schematic in Figure~\ref{fig:distances} and being $x$ and $y$ the horizontal and vertical directions and $\left\lbrace \mathbf{e}_x, \mathbf{e}_y \right\rbrace$ the standard basis)
		 \begin{itemize}
 	 		\item $\bm{F}^{K2,I-J}_\text{G.S.} = -0.4831 \, \mathbf{e}_x - 0.2789 \, \mathbf{e}_y \quad$ eV / \AA
 	 		\item $\bm{F}^{K3,I-J}_\text{G.S.} = +0.4831 \, \mathbf{e}_x - 0.2789 \, \mathbf{e}_y \quad$ eV / \AA
 	 		\item $\bm{F}^{I,I-J}_\text{G.S.} = -0.9661 \, \mathbf{e}_x \quad$ eV / \AA
 	 		\item $\bm{F}^{J,I-J}_\text{G.S.} = +0.9661 \, \mathbf{e}_x \quad$ eV / \AA 	 		
 	 		\item $\bm{F}^{K1,I-J}_\text{G.S.} = -0.4831 \, \mathbf{e}_x + 0.2789 \, \mathbf{e}_y \quad$ eV / \AA 	 	
 	 		\item $\bm{F}^{K4,I-J}_\text{G.S.} = +0.4831 \, \mathbf{e}_x + 0.2789 \, \mathbf{e}_y \quad$ eV / \AA 	 		
		 \end{itemize}
	\end{itemize}
	\item General description of our hybrid MD implementation:
 	\begin{itemize}
		\item Time integration of~(\ref{eq:newton2}): Velocity-Verlet scheme for the Hamiltonian equations and exact integration of the Ornstein-Uhlenbeck process~\cite{BouRabee10}:
		 \begin{itemize}
 	 		\item $\bm{v}^{(n+1/2)}=\bm{v}^{(n)}+\frac{\bm{F}^{(n)}}{2 m} \delta t$	 		
 	 		\item $\bm{r}^{(n+1)}=\bm{r}^{(n)}+\bm{v}^{(n+1/2)} \delta t$	
 	 		\item $\bm{F}^{(n+1)}= - \nabla E_{(\bm{r}^{I(n+1)})}$
 	 		\item $\bm{v}^{(n+1)*}=\bm{v}^{(n+1/2)}+\frac{\bm{F}^{(n+1/2)}}{2 m} \delta t$ 	 		 	 		
 	 		\item $\bm{v}^{(n+1)}=\alpha \bm{v}^{(n+1)*}+ \sqrt{\frac{\text{k}_\text{B} T}{m} (1-\alpha^2)} \bm{R}$
 	 		
 	 		where $\alpha = \exp{\left( -\gamma \delta t \right)}$ and $\bm{R}$ is a vector whose random components that are mutually uncorrelated, Gaussian, and have zero-mean and unit variance. 	
		 \end{itemize}		
		\item Neighbor list: Cell structures and linked cells method (see~\cite{Allen})
	\end{itemize}
	\item Settings:
 	\begin{itemize}
		\item Timestep: $\delta t = 5. 10^{-4}$ ps
		\item Temperature of thermostat: $ T = 300$ K
		\item Langevin damping coefficient $\gamma = 1.0$ ps$^{-1}$
		\item Number of steps for thermalization: $10^4$
		\item Deformation rate $0.5$ \% / ps
		\item Box deformation performed every step
		\item Computation of the criterion / substitution performed every step		
		\item Neighbor list updated every $100$ steps		
	\end{itemize}

\end{itemize}
}




\bibliographystyle{elsarticle-num} 
\bibliography{manuscript}

\end{document}